# Scalable, Non-contact Determination of Electric Properties of Nanostructures via Electro-Rotation in Water Solution

Yun Huang[1], Kai Xu[2], Zexi Liang[1], Huaizhi Li[1], Wenjuan Zhu[2], Donglei Emma Fan[1,3,4*]

[1]Materials Science and Engineering Program, Texas Materials Institute, University of Texas at Austin, TX 78712, USA

[2]Electrical and Computer Engineering, University of Illinois at Urbana-Champaign, Urbana, IL 61801, USA

[3]Walker Department of Mechanical Engineering, University of Texas at Austin, Austin, TX 78712, USA

[4]Chandra Family Department of Electrical and Computer Engineering, The University of Texas at Austin, Austin, TX, USA

Correspondence should be addressed to: dfan@austin.utexas.edu

## ABSTRACT

Breakthroughs in nanotechnology have enabled the large-scale fabrication of nanoparticles with varied compositions and structures. Yet, evaluating their electrical conductivities remains challenging due to high volume and individual variability. We report a rapid, non-contact, and parallel method to characterize longitudinal nanostructures, including insulators, semiconductors, and conducting metal oxides by using $MoO_3$, $MoS_2/MoO_2$, and $MoS_2$ nanoribbons, produced at different fabrication stages, as a model system. Leveraging our semi-quantitative model based on Maxwell-Wagner and electrical double-layer polarization, electric conductivities of various nanoparticles are determined from their distinct electro-rotation behaviors in water, spanning six orders of magnitude. The results agree well with standard four-probe measurements. The technique, measuring multiple nanoparticles at once, without the use of electrical contact, can be easily scaled up for parallel determination of particles' electric conductivities. These findings highlight a non-destructive, rapid, and simple characterization method promising to bring nanomaterials closer to practical applications in electronics, optics, sensing, catalysis, and robotics.

Nanostructure-enabled devices are impacting modern society in diverse fields, including sensors,[1] robotics,[2] energy,[3] medicine,[4,5] and electronics.[6,7] Their performance and reliability highly depend on the physical and chemical properties of integrated nanomaterials. However, due to the ultrasmall dimensions and local variations of fabrication conditions, nanostructures are often grown with intrinsic heterogenies,[8, 9] even if made using the same procedure. For instance, Au nanoparticles show remarkable catalytic heterogeneity.[10] Photoluminescence spectrum from individual CdSe nanowires can yield a shift of 62 meV.[11] Tenacity of carbon fibers can vary by 25%.[12] Therefore, rapid, non-destructive, and high-throughput property assessment is essential for nanomaterial selection and application.

The straightforward method for measuring materials' electric conductivities using electrode pairs requires careful selection of contact materials, where Fermi level pining effects can be significant.[13, 14] Although four-probe measurement (FPM) effectively eliminates contact resistance,[15, 16] its laborious fabrication and measurement greatly restrict its use for profiling nanoparticles. Non-contact techniques like Kelvin probe and scanning photocurrent microscopy can map surface potential and carrier distribution of nanostructures,[17-20] yet they are time-consuming and expensive. These challenges demand a fast, inexpensive, and reliable technology to characterize electronic nanomaterials.

Electrokinetic techniques such as dielectrophoresis, electrorotation, and electro-orientation have been developed to manipulate biological and synthesized micro/nanoparticles.[21, 22] Dielectrophoresis can also be used to charaterize the biophysical properties of cells.[23] In 2012, wireless electrical manipulation was firstly proposed to probe electronic properties of micro/nanoparticles.[24] Experimentally, the mechanical behaviors of nanoparticles in water solution become complex at AC electric fields below 100 kHz due to the interaction of electric double layers with the polarized particles. Leveraging a non-aqueous medium, such as mineral oil or dipropylene glycol (DPG), Akin, Yi, Feldman, *et al.* quantified the electrical conductivity of Si nanowires by electro-orientation spectroscopy (EOS).[25, 26] The method, however, requires strong *E*-fields and is restricted to particles with a conductivity less than 10 S/m due to the much lowered electric manipulation efficiency and high resistivity of the non-aqueous media. Thus, a broadly applicable method for determining electronic properties with a wide range remains lacking. Electrorotation spectroscopy (ERS) based on non-synchronous rotation of nanoparticles in a high-frequency electric field in water is particularly desirable owing to its simplicity in sample handling, high efficiency, and accuracy (**SI Note 2**).

Recently, we established a semi-quantitative model to analyze the electrokinetic behavior of nanoparticles in water across kHz to MHz AC frequencies. The model, built on Maxwell-Wagner theory and experimentally determined electrical-double-layer polarization, permits quantitative calculation of electrokinetic effects on nanoparticles in aqueous solutions.[27, 28] Combining this model with experimental manipulation results, we firstly report the unravelling of electric properties of nanostructures based on their electrorotation behaviors in aqueous solutions, including insulators ($MoO_3$), semiconductors ($MoS_2$), and conducting hybrids ($MoO_2/MoS_2$), spanning six orders of magnitude in electric conductivity. The technique is non-contact, scalable, and fast, enabling simultaneous measurements of multiple nanoparticles within minutes. It can reliably characterize materials with conductivities up to $10^2$ S/m, extending at least one order of magnitude beyond EOS's working range, with easy processing of nanoparticle dispersion in water. Importantly, the results agree with standard four-probe measurements, typically within one order of magnitude, validating the reported approach.

Three types of nanoribbons made of $MoO_3$, $MoS_2/MoO_2$ hybrid, and $MoS_2$ are synthesized at different stages during a two-step sulfurization process.[29] In short, $MoO_3$ nanoribbons are grown by hydrothermal synthesis, then sulfurized via chemical vapor deposition at 500 °C to form $MoS_2/MoO_2$ hybrids. Further sulfurization at 900 °C yields high-purity $MoS_2$ nanoribbons. Each step produces large quantities of nanoribbons with well controlled longitudinal morphologies as shown in the scanning electron microscope (SEM) images (**Figure 1a-c)**. Most nanoribbons are several micrometers long. After sulfurization, the $MoS_2/MoO_2$ hybrid (**Figure 1e**) and $MoS_2$ nanoribbons (**Figure 1f**) show slight curvature compared to the original $MoO_3$ templates (**Figure 1d**). The nanoribbon widths are 280 ± 76 nm, 230 ± 81 nm, and 250 ± 91 nm for $MoO_3$, $MoS_2/MoO_2$ hybrid, and $MoS_2$ respectively (**Figure 1g-i**). The thickness increases through the fabrication process, from 90.7 ± 24.4 nm for $MoO_3$, 119.5 ± 26.5 nm for $MoS_2/MoO_2$ hybrid, to 171.1 ± 45.7 nm for $MoS_2$. The representative atomic force microscopy characterizations are provided in **Figure S1**.

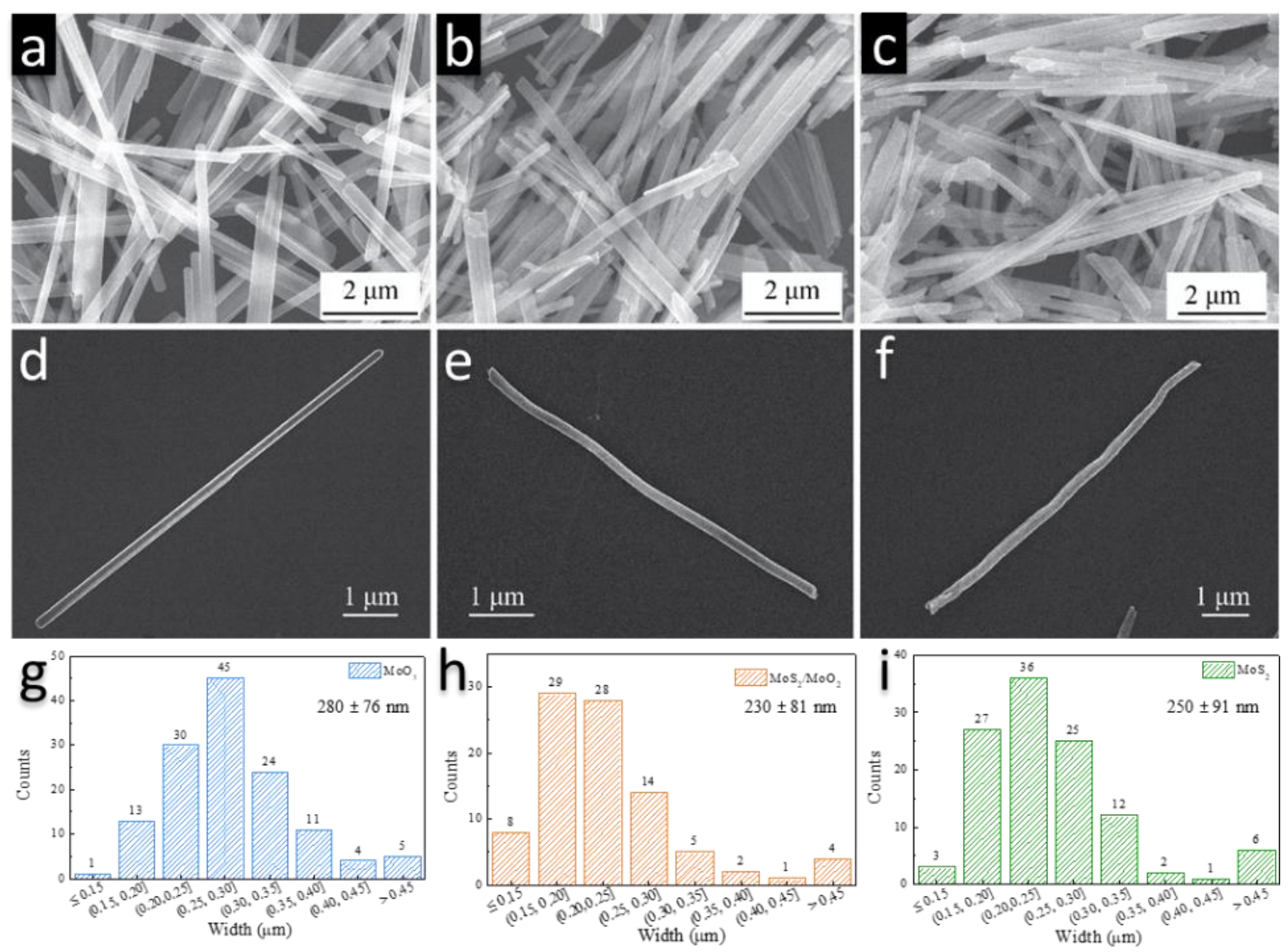

**Figure 1.** *Nanoribbons made of $MoO_3$, $MoS_2/MoO_2$ and $MoS_2$. SEM of massive (a) $MoO_3$, (b) $MoS_2/MoO_2$ hybrid, and (c) $MoS_2$ nanoribbons, and (d-e) typical individual $MoO_3$, $MoS_2/MoO_2$ and $MoS_2$ nanoribbons, respectively. (g-i) Distribution of Nanoribbons' widths.*

$MoS_2$ nanoribbons can be efficiently manipulated by using combined DC and AC electric fields.[29] In addition to controlled transport, they can rotate and align.[23, 27, 28] Here, we rotate the three distinct types of nanoribbons and determine their electric conductivities from ERS. The experimental setup consists of a quadrupole microelectrode enclosing a 500 μm × 500 μm area at the center as shown in **Figure 2a**. In a typical test, a droplet of nanoribbon suspension (20 μl) is dispersed on the microelectrodes and settled for 2 mins, before a rotating electric field is created by four AC voltages with sequential 90° phase shift. For convenience, rotation is programmed in the clockwise direction. Nanoribbon motion is recorded by a CCD-equipped optical microscope. **Figure 2b** shows snapshots of a $MoS_2$ ribbon rotating at 10 kHz (counterclockwise) and 300 kHz (clockwise). Notably, the three materials exhibit distinct electro-rotation behaviors (**Movie 1-3**).

For each type of nanoribbon, we measure the rotation speed across 5 kHz to 1.0 MHz under a constant AC-field magnitude (**Figure 2c-e**). $MoO_3$ nanoribbons always display co-field rotation with the lowest overall speed. $MoS_2/MoO_2$ nanoribbons rotate counter to the field from 5 kHz to 100 kHz and stop above 200 kHz. Interestingly, $MoS_2$ nanoribbons reverse their rotation direction from counter-field to co-field as the frequency increases near 50 kHz. The measurement is well reproducible as evidenced by the narrow error bars from repeated measurements (3–5 times) on the same nanoribbon.

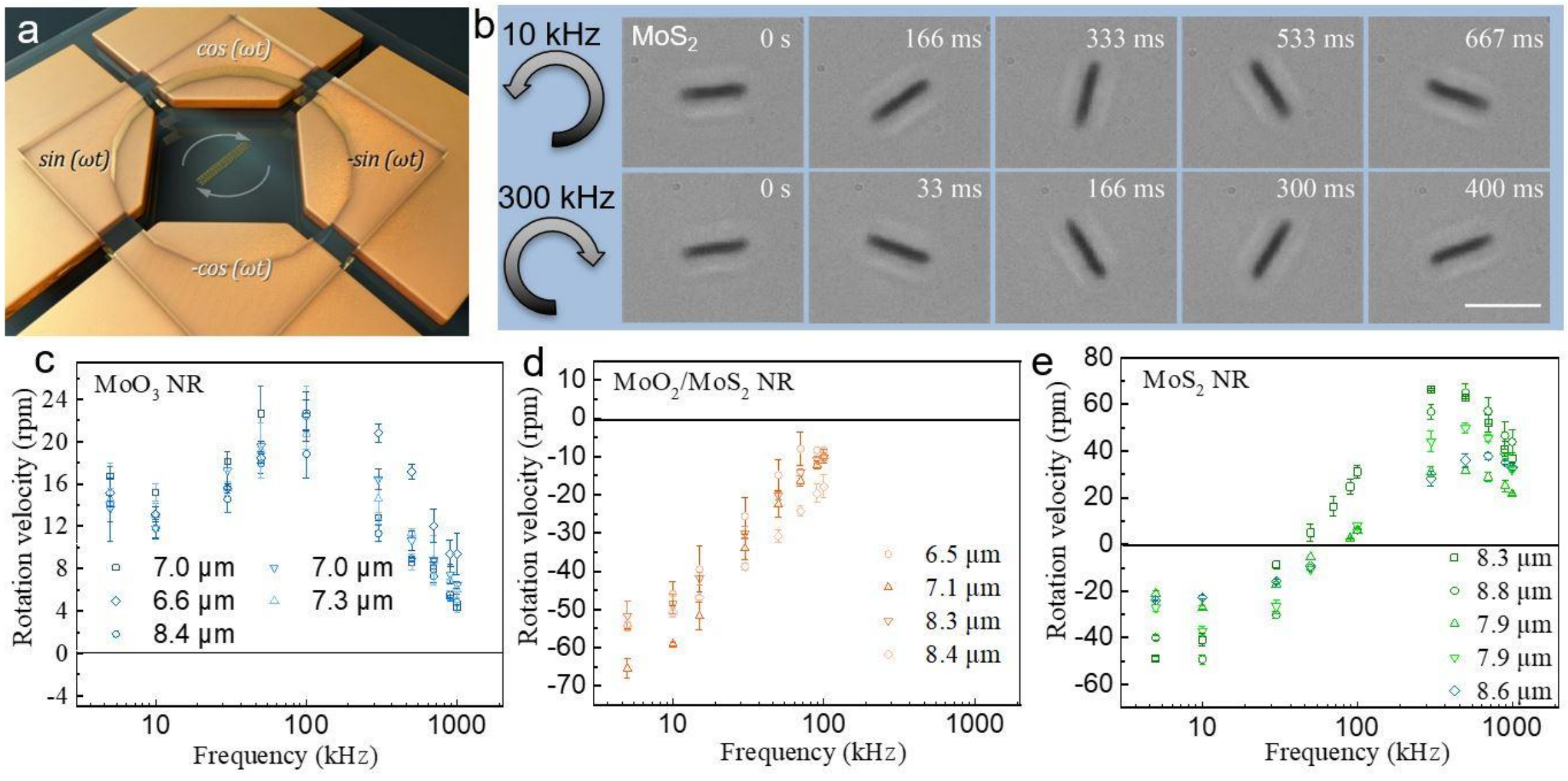

**Figure 2.** *Frequency dependent electro-rotation of $MoO_3$, $MoS_2/MoO_2$, $MoS_2$ nanoribbons. (a) Schematic of experimental setup. (b) Sequential snapshots of a rotating $MoS_2$ nanoribbon showing counter-field rotation and co-field rotation at 10 $V_{pp}$ (peak-to-peak), 10 kHz and 300 kHz respectively. Scale bar: 10 µm.(c) ERS of $MoO_3$ (20 $V_{pp}$), (d) $MoS_2/MoO_2$ (10 $V_{pp}$), and (e) $MoS_2$ (10 $V_{pp}$) nanoribbons.*

The observed rotation behaviors align with previous findings for nanowires of similar electronic types. For instance, the counter-field rotation of $MoS_2/MoO_2$ nanoribbons matches that of metallic Au and Pt nanowires,[24, 30] reflecting the metallic nature of $MoO_2$. Similarly, the frequency-dependent rotation chirality of $MoO_3$ and pure $MoS_2$ nanoribbons aligns with those of insulating $SiO_2$ and semiconducting Si nanowires, respectively.[24, 27, 30, 31] Given their similar dimensions, the distinct rotation behaviors are qualitatively attributed to their electronic properties. However, quantitatively determining electric conductivities from the ERS obtained in water is challenging due to the complex effect of the electric double layer below 100 kHz. Although using low-dielectric organic media can mitigate this issue, it reduces efficiency, measurable accuracy and range. To overcome this, we employ our semi-quantitative model based on Maxwell-Wagner theory and experimentally determined electrical double-layer polarization.[27]

In a rotating $E$-field $E(t) = E_0 Re[(\hat{\boldsymbol{x}} - i\hat{\boldsymbol{y}})e^{i\omega t}]$, where the frequency is $\omega$ and electric-field strength is $E_0$, a nanoribbon is polarized with an effective dipole moment $\boldsymbol{p}$, which interacts with the $E$-field, resulting in a time-averaged electric torque given by:

$$\langle \boldsymbol{\tau_e} \rangle = -\frac{1}{2} E_0 \mathrm{Im}(\underline{\boldsymbol{p}})\hat{\boldsymbol{z}}, \tag{1}$$

where $\underline{\boldsymbol{p}}$ is the phasor representation of the dipole moment $\underline{\boldsymbol{p}}(t) = \boldsymbol{p}\exp(i\omega t)$. In aqueous solution, nanoribbons (length: tens of micrometers; moving speed: a few body length per second) reside in low Reynold's number regime, where viscous drag instantly balances electric torque, given by:

$$\boldsymbol{\tau}_{\eta} = \gamma \frac{d\theta}{dt} = \boldsymbol{\tau_e}. \tag{2}$$

Here, $\gamma$ is rotational drag coefficient, which is a function of a particle's geometry and suspension medium's viscosity (see **Note 1**), and $\frac{d\theta}{dt}$ is rotational velocity. The instant balance is confirmed in experiments. As shown in **Movies 1-3**, the rotation of a nanoribbon instantly reaches terminal velocity, which linearly increases with applied electric torque ($\boldsymbol{\tau_e}$), agreeing with **Eq.2**. Combining **Equations 1-2**, we can readily calculate $\mathrm{Im}(\underline{\boldsymbol{p}})$, the imaginary part of the effective dipole moment from the experimentally measured ERS in **Figure 2 (c-e)**.

To unravel $\mathrm{Im}(\underline{\boldsymbol{p}})$ and its correlation with the electrical properties of nanoparticles, we model a nanoribbon as a prolate spheroid. Its dipole moment includes two components, the classic Maxwell-Wagner (MW) relaxation, owing to the distinct permittivity of a particle and its suspension medium at the interface,[32] and the electrical double layer (EDL) formed in response to the interfacial charge introduced by MW relaxation given in the following equation:[27, 33]

$$\underline{\boldsymbol{p}} = \underline{\boldsymbol{p}_{MW}} + \underline{\boldsymbol{p}_{EDL}} \,. \tag{3}$$

The dipole moment ($\boldsymbol{p}_{MW,n}$) from the MW relaxation in an electric field ($E_n$) along each principal axis $n$ (x, y, z respectively) of the nanoribbon can be expressed as:[32]

$$\boldsymbol{p}_{MW,n} = 4\pi a_x a_y a_z \varepsilon_m \widetilde{K}_n \boldsymbol{E}_n \tag{4}$$

where $a_x$, $a_y$, and $a_z$ are the length of three semi-axes, respectively, $\widetilde{K}_n$ is the Clausius-Mossotti factor that governs the MW polarizability, which equals:

$$\widetilde{K}_n = \frac{\varepsilon_p^* - \varepsilon_m^*}{\varepsilon_m^* + L_n(\varepsilon_p^* - \varepsilon_m^*)} \tag{5}$$

where $\varepsilon_p^* = \varepsilon_p - i\sigma_p/\omega$ and $\varepsilon_m^* = \varepsilon_m - i\sigma_m/\omega$ are the complex permittivity of the particle and suspension medium, respectively, and $L_n$ is the depolarization factor which depends on the geometry of the nanoparticles. By combining **equations 4 and 5**, $\widetilde{K}_n$, and therefore $\boldsymbol{p}_{MW,n}$, can be readily calculated from the permittivity ($\varepsilon_p$, $\varepsilon_m$) and conductivity ($\sigma_p$, $\sigma_m$) of both the particle and medium, the frequency of the rotating *E*-field ($\omega$), and the geometry-governed $L_n$.

Furthermore, due to the large aspect ratio, nanoribbons exhibit a much stronger dipole moment along the long axis (y) compared to the other two axes. As shown in **Figure S2**, the calculated $Im(\widetilde{K}_n)$ values along the width and thickness of a $MoS_2$ nanoribbon are < 0.04% of that along the length, 1 kHz to 1 MHz. Meanwhile, the dipole component along the thickness direction is perpendicular to the rotation plane and thus does not contribute to the rotational torque. Therefore, the effective dipole moment for generating electric torque can be approximated from the dominant component, $\boldsymbol{p}_{MW,y}$.

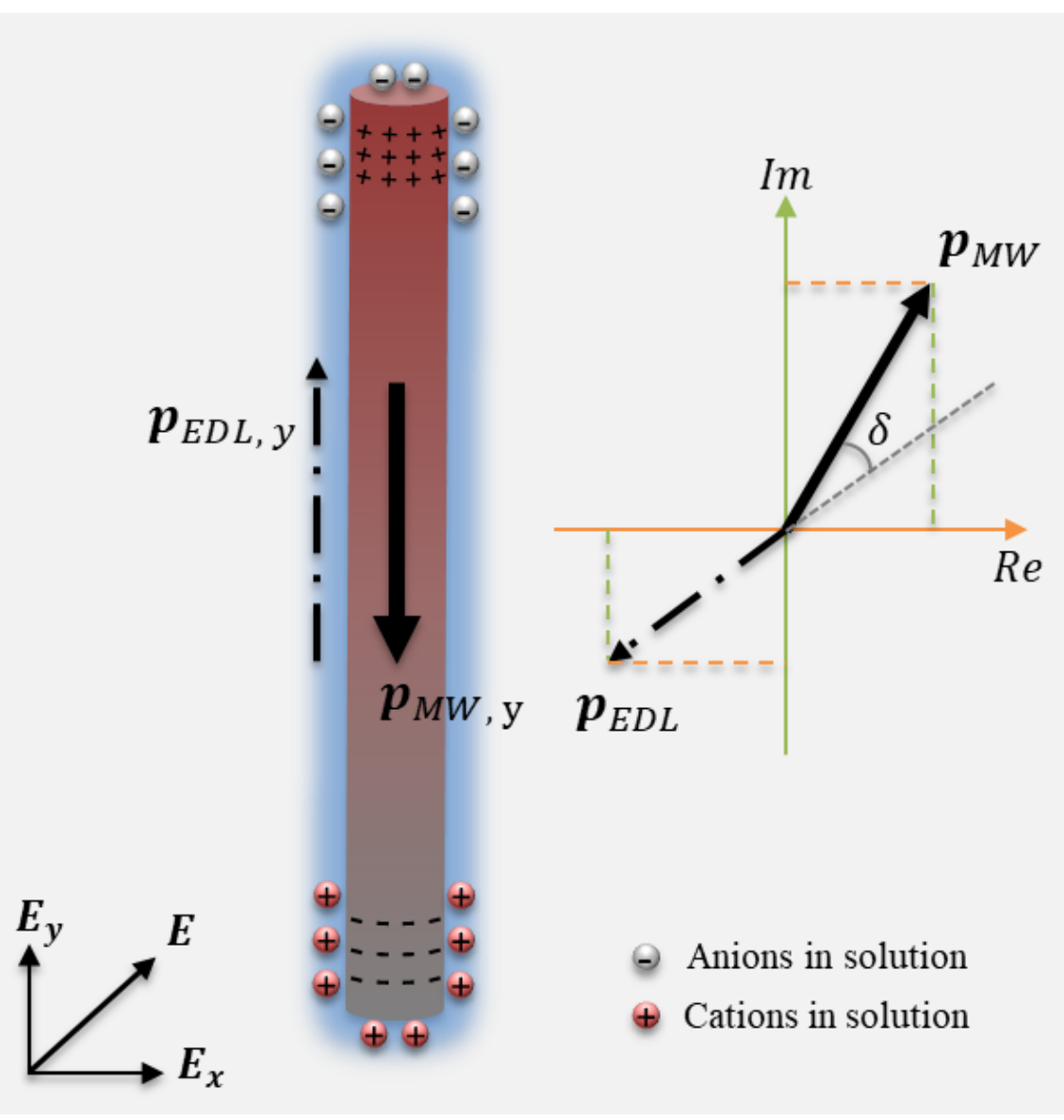

**Figure 3.** *Model based on MW relaxation* and *EDL for the calculation of electro-rotation spectrum of a longitudinal nanostructure.*

Next, we calculate the dipole moment induced by EDL. In a solution containing charged ions, MW polarization of the nanoparticle induces an asymmetric distribution of charges in the surrounding diffuse layer (**Figure 3**), forming the EDL. The formation of EDL depends on the MW dipole ($\boldsymbol{p}_{MW}$), AC $E$-field frequency, diffusion speed, and concentration of charged ions (or molecules) in suspension. Treating the EDL as an equivalent resistor-capacitor (RC) circuit with a time constant $\tau_{RC}$, the imaginary part of the complex EDL-induced dipole moment $\mathrm{Im}(\underline{\boldsymbol{p}_{EDL}})$ is given by:[27]

$$\mathrm{Im}(\boldsymbol{p}_{EDL}) = -[\mathrm{Re}(\boldsymbol{p}_{MW}) \sin\delta + \mathrm{Im}(\boldsymbol{p}_{MW}) \cos\delta] \frac{1}{\sqrt{\omega^2 \tau_{RC}^2 + 1}}, \tag{6}$$

where $\delta$ is the phase lag between $\boldsymbol{p}_{EDL}$ and $\boldsymbol{p}_{MW}$, and $\tan\delta = -\omega\tau_{RC}$. Combining **equations (1) to (6)**, we obtain frequency-dependent electric torque on a longitudinal nanoparticle in water.

By fitting the ERS spectra using this quantitative model, we determine the electrical conductivities of $MoO_3$ and $MoS_2$ nanoribbons. The parameters used in the calculations are listed in **Table S1** and detailed in SI **Note-1**. Specifically, the conductivity of water ($6 \times 10^{-4}$ S/m) is measured experimentally, and a relative permittivity of 80 is used. The time constant ($\tau_{RC}$) is extracted from fitting ERS data of Au nanowires **(Figure S3)**.

**Figure 4** presents the rotation speed spectra of several nanoribbons with similar geometries, along with theoretical ERS spectra across a range of electrical conductivities. Clearly, the ERS response is highly sensitive to electric conductivity. **Figure 4a and 4b** show that even a small change (less than one order of magnitude) causes an evident shift in the ERS spectrum. By closely matching the experimental and theoretical curves, we determine the conductivities of $MoO_3$ and $MoS_2$ nanoribbons to be $3.5 \times 10^{-4}$ and 0.6 S/m, respectively, consistent with their electronic properties—$MoO_3$ with a large bandgap of ~3.2 eV is insulating,[34, 35] while $MoS_2$ with a bandgap of 1.2 - 1.9 eV is a semiconductor.[36]

To confirm the reproducibility, we measure 10 additional $MoO_3$ and $MoS_2$ nanoribbons, as shown in **Figure S4** and **S5**, respectively. The obtained electric conductivity shows excellent agreement within each type with electric conductivities ranging from $3.5 \times 10^{-4}$ to 1 S/m. These studies prove the feasibility of ERS for rapid electric characterization of nanomaterials with a wide conductivity range. However, for $MoS_2/MoO_2$ nanoribbons, the ERS fitting method cannot precisely determine electrical conductivity (**Figure 4c**) due to largely overlapping theoretical fittings when conductivities are > 100 S/m. Nevertheless, the result still unambiguously supports that $MoS_2/MoO_2$ nanoribbons are electrically conducting, which is consistent with the presence of conductive $MoO_2$ in the hybrids.

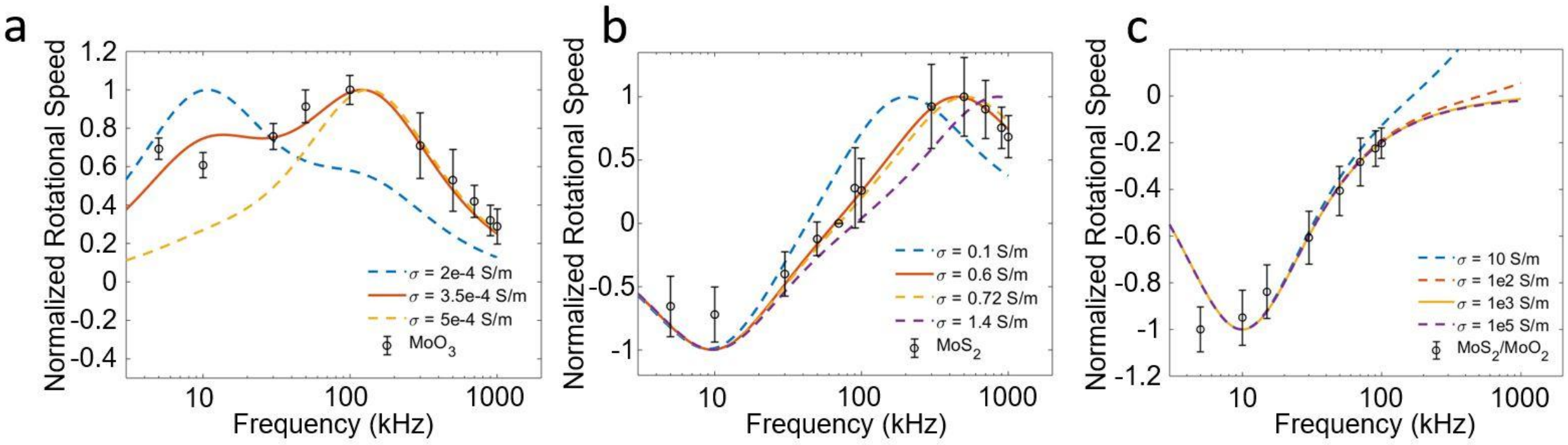

**Figure 4.** Normalized electro-rotation spectra with corresponding model fitting curves for determining electric conductivities of *(a)* $MoO_3$ *(b)* $MoS_2$ *and (c)* $MoS_2/MoO_2$ *hybrid nanoribbons.*

In our study, accurately measuring the width and thickness of individual nanoribbons for ERS model fitting is a challenge. As shown in **Figure 1** and **Figure S1**, these dimensions vary across different ribbons. For the assessment, we investigate how changes in width and thickness may alter the theoretical fitting curves. Interestingly, the extracted conductivity of $MoO_3$ nanoribbons remains nearly unchanged— even when we double or halve the average width and thickness (**Figure S6**). $MoS_2$ nanoribbons show similar behaviors (**Figure S7**). When the width or thickness of a nanoribbon is doubled relative to the average, the extracted electrical conductivity changes only slightly. This indicates that ERS spectra are primarily sensitive to a nanoribbon's electrical conductivity and length, where the latter is measurable via optical microscopy. These findings further support ERS as a rapid, non-contact technique for quantitatively characterizing electrical conductivity across a large range—from insulators to semiconductors—while also enabling unambiguous identification of metallic nanoparticles.

Finally, we assess the accuracy of the ERS-determined electrical conductivity by comparing it with results from standard four-probe measurements, which offers precise electronic characterization by eliminating interfacial contact resistances.[37] We begin FPM by dispersing nanoribbons onto a thermally oxidized Si substrate with a 100-nm $SiO_2$ coating. Next, four microelectrodes are defined by e-beam lithography, followed by the electron-beam deposition of 50/100 nm Ti/Au. **Figure 5a** shows a typical $MoS_2$ nanoribbon device. While a constant current ($I$) is applied between the V1 and V4 electrode pads, the electrical potential between V2 and V3 electrodes is measured, and the difference is taken as $\Delta V$. The value of the resistance between the V2 and V3 electrode ($R$) can be calculated accurately using $R = \Delta V / I$ ,where any contact resistance or difference between electrode contacts can be eliminated.[37] The electric conductivity is determined using both the known distance between V2 and V3 and the cross-section area of the nanoribbon. For each type of ribbon, at least 5 individual devices were measured (full data in **Figure S5**). **Figure 5b** plots the typical resistance of a single $MoO_3$, $MoS_2/MoO_2$ and $MoS_2$ nanoribbon, respectively. The electric resistances of the three materials span several orders of magnitudes, with $MoO_3$ being highly insulating and $MoS_2/MoO_2$ being conductive. The electric conductivities of $MoS_2$ and $MoO_3$ show a narrow distribution (green and blue panels in **Figure 5b**), while the value of $MoS_2/MoO_2$ nanoribbons varies in a larger range (light red panel). These results agree with the high purity of obtained $MoO_3$ and $MoS_2$,[29] while, $MoS_2/MoO_2$ nanoribbons are metal/semiconductor hybrids whose electrical conductivity can change significantly, depending on the two-component internal distribution.[29]

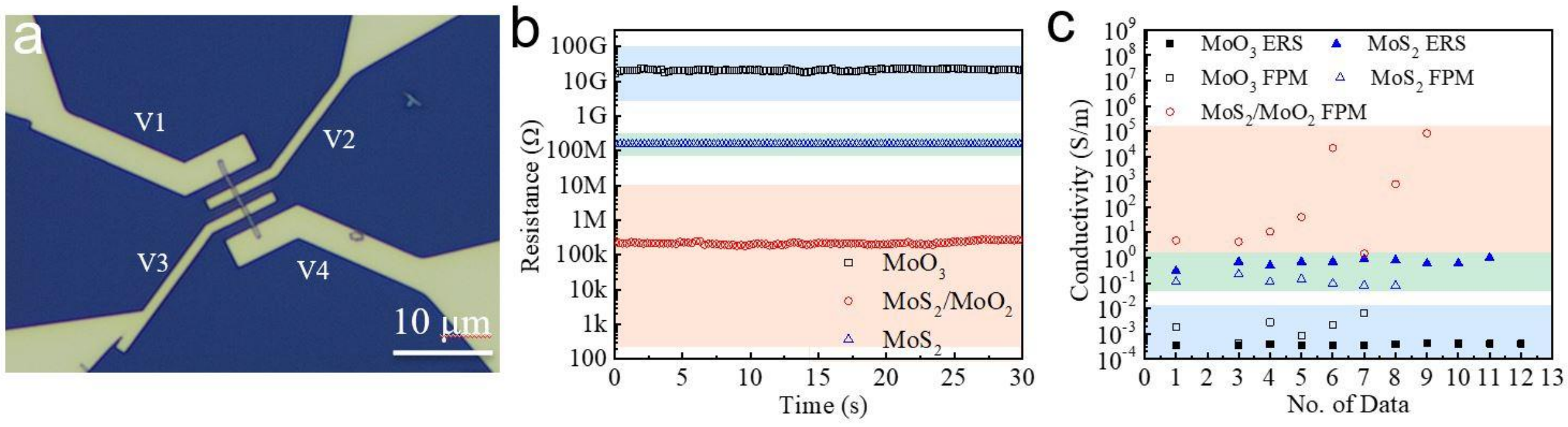

**Figure 5.** *Comparison of electric conductivity determined from four-probe measurement and electro-rotation. (a) Optical microscope image of a four-probe characterization device made on a $MoS_2$ nanoribbon. (b) Typical electric resistances of $MoO_3$, $MoS_2/MoO_2$ and $MoS_2$ nanoribbons measured by FPM. (c) Conductivities of nanoribbons determined by FPM (open symbols) agree well with those obtained from electro-rotation spectra (solid symbols) for $MoO_3$ and $MoS_2$.*

Next, we compare the FPM results with those obtained from ERS, focusing on $MoO_3$ and $MoS_2$ nanoribbons whose electrical conductivities fall within the quantitative working range of the ERS method. The ERS-fit electrical conductivities (**Figure S4** and **Figure S5**) and those measured by FPM are plotted in **Figure 5c**. The results show excellent agreement, all within the same order of magnitude. The minor differences could be attributed to the inherent variation of nanoparticles, uncertainties of the thickness and width employed in ERS modeling (**Figure S6** and **Figure S7**), and potentially altered surface chemistry and conductance of $MoS_2$ and $MoO_3$ in water solutions.

In the frequency range of 5 kHz to 1 MHz, the electrical conductivities of particles can be regarded as constant and equivalent to their DC values, since frequency-dependent conductivity is observed typically at much higher frequencies. For instance, in $MoS_2$, notable frequency-dependent behavior emerges in the terahertz region.[38, 39] Indeed, the ERS technique in this work determines static electric conductivity of a particle, same as the four-probe method.

Light illumination may also increase the electric conductivity of semiconductors due to photo-generated electron-hole pairs.[27, 28] For instance, our previous work shows the photocurrent of synthesized $MoS_2$ increases ~20% under the illumination of 700 nm laser with an intensity of $\sim 5 \times 10^5$ W/m$^2$.[29] In this work, however, such an effect is minimal owing to our careful control of light illumination in the ERS test (~16, 000 lux), where the electric resistances measured by FPM for $MoS_2$ and $MoO_3$ nanoribbons remain the same in both dark and light conditions (~300, 000 lux) (**Figure S9**).

Collectively, our work clearly demonstrates multiple advantages of the ERS method for evaluating the electronic properties of nanoparticles. Despite uncertainties in the thickness and width of individual nanoribbons, the conductivities obtained by using ERS remain within one order of magnitude of those measured by the standard four-probe method, highlighting its quantitative reliability. ERS is also adaptable to nanostructures with diverse geometries, and its measurement accuracy can be further improved by controlling particle shape during fabrication. Overall, the ability to determine the electrical conductivity of unknown micro- and nanoscale particles within minutes in a non-contact, non-destructive manner represents a major breakthrough over conventional methods; one that could significantly accelerate the practical application of nanoparticles in optoelectronics, sensing, energy systems, and micro/nanorobotics.

In conclusion, we report a non-contact, parallel, and high-speed method for quantitative characterization of electrical conductivities of longitudinal nanoparticles. A model system made of $MoO_3$, $MoS_2$, and $MoS_2/MoO_2$ nanoribbons created at different fabrication stages is examined. Simply by electro-rotating the nanoribbons at different electric frequencies and conducting spectrum fitting using our theoretical model, we can readily extract electric conductivity of $MoO_3$, $MoS_2$ nanoribbons and evidently determine the metallic nature of $MoS_2/MoO_2$ nanoribbon. The theoretical model considers both nanoparticles' MW relaxation and EDL polarization, which enables the ERS spectrum fitting for nanoparticles rotating in aqueous solution. The results well agree with the standard four-probe measurements. Notably, our method only takes a few minutes to determine multiple nanoribbons' electric conductivities without any elaborate instrumentation. The all–electric-field-based device scheme and instrumentation make the ERS technique highly compatible with current lab-on-a-chip systems for sorting and manipulation, such as dielectrophoresis[40-42] (DEP) and traveling-wave tweezers[43-46], thereby addressing a critical gap in current nanoparticle sorting technologies. Distinct from the previous reported method where special dispersion solutions are selected to minimize the EDL effect, our work can be readily extended to nanoparticles in different solutions. We expect this work to accelerate the practical applications of nanoparticles, impacting optoelectronics, biomedical sensing, energy conversion and storage, and micro/nanorobotics.

## Associate Content

Supporting information. Additional information, including supporting movies and figures, calculation parameters and AFM images of the nanoribbons are provided. The effects of parameters such as width and thickness on the extraction of electrical conductivity are discussed. A detailed comparison between ERS and EOS is presented.

## Data Availability

The data that supports the findings of this study are available within the article and its supplementary material.

## Author Information

### Corresponding Author

**Donglei Emma Fan** – Materials Science and Engineering Program, Texas Materials Institute, Walker Department of Mechanical Engineering, Chandra Family Department of Electrical and Computer Engineering, University of Texas at Austin, Austin, TX 78712, USA

Email: dfan@austin.utexas.edu.

### Notes

The authors declare no competing financial interest.

## Acknowledgements

This research is supported by National Science Foundation (EECS-1710922, EECS-1930649, and PFI 2219221), NIH (R43ES037145), and the Welch Foundation (F-1734) in part.